\documentclass{article}

\usepackage[dvips]{graphics,graphicx,color}
\usepackage{a4,indentfirst,latexsym,amsfonts,amsmath,amssymb,array}
\usepackage{wrapfig}

\begin{document}

\title{Radiation defects in silicon due to hadrons and leptons, their annealing and 
influence on detector performance}

\author {Ionel Lazanu\footnote{University of Bucharest, POBox MG-11, Bucharest-Magurele, 
Romania} \hspace{2 pt}
and Sorina Lazanu\footnote{National Institute for Materials Physics, POBox MG-7, 
Bucharest-Magurele, Romania}}

\date{\today}
\maketitle
\begin{abstract}
A phenomenological model was developed to explain quantitatively, without free parameters, 
the production of primary defects in silicon after particle irradiation, the kinetics of 
their evolution toward equilibrium and their influence on detector parameters. The type of 
the projectile particle and its energy is considered in the evaluation of the 
concentration of primary defects. Vacancy-interstitial annihilation, interstitial 
migration to sinks, vacancy - impurity complexes ($VP$, $VO$, $V_2O$), and divacancy 
($V_2$) formation are taken into account in different irradiation conditions, for 
different concentrations of impurities in the semiconductor material, for 20 and 0 $^o$C. 
The model can be extended to include other vacancy and interstitial complexes. The density 
of the reverse current in the detector after irradiation is estimated. Comparison with 
experimental measurements is performed. A special application considered in the paper is 
the modelled case of the behaviour of silicon detectors operating in the pion field 
estimated for the LHC accelerator, under continuum generation and annealing.

\medskip
\textbf{PACS}: \\
29: Experimental methods and instrumentation for elementary-particle and nuclear physics\\
81: Materials science\\
78: Optical properties, condensed-matter spectroscopy and other interactions of radiation 
and particles with condensed matter

\medskip

\end{abstract}

\section{Introduction}

The use of silicon detectors in high radiation environments, as to be expected in future 
high energy accelerators, poses severe problems due to changes in the properties of the 
material, and consequently influences the performances of detectors.

The incident particle, hadron or lepton, interacts with the electrons and with the nuclei 
of the semiconductor lattice. It losses its energy in several processes, which depend on 
the nature of the particle and on its energy. The effect of the interaction of the 
incident particle with the target atomic electrons is ionisation, and the characteristic 
quantity for this process is the energy loss or stopping power. The nuclear interaction 
between the incident particle and the lattice nuclei produces bulk defects and this 
phenomenon is studied in the present paper. As a result of this interaction, if the 
primary projectile is a particle, one or more light particles are formed, and usually one 
(or more) heavy recoil nuclei. This recoil nucleus has charge and mass numbers equal or 
lower than that of the medium. After this first interaction, the recoil nucleus or nuclei 
are displaced from the lattice positions into interstitials. Then, the primary knock-on 
nucleus, if its energy is large enough, can produce the displacement of a new nucleus, and 
the process continues as long as the energy of the colliding nucleus is higher than the 
threshold for atomic displacements. We denote these displacement defects, vacancies and 
interstitials, as primary defects, prior to any further rearrangement. 
In silicon these defects are essentially unstable and interact via migration, 
recombination, annihilation or produce other defects.

As a consequence of the degradation to radiation  of the semiconductor material, an 
increase of the reverse current due the reduction of the minority carrier lifetime, a 
reduction of the charge collection efficiency and a modification of the effective doping, 
due to the generation of trapping centres, are observed in the detector characteristics.
In this paper, for the first time, a phenomenological model was developed to explain 
quantitatively, without free parameters, the mechanisms of production of the primary 
defects during particle irradiation, the kinetics of their evolution toward stable defects 
and equilibrium and the influence of the defects on detector parameters. The effects of 
the incident particle type, of its kinetic energy and of the irradiation conditions on the 
concentration of defects are studied. Vacancy-interstitial annihilation, interstitial 
migration to sinks, vacancy - impurity complexes ($VP$, $VO$ and $V_2O$), and divacancy 
($V_2$) formation are considered in different irradiation conditions, for different 
concentrations of impurities in the semiconductor material and at different temperatures 
near room temperature. The model can be extended directly to include the effects of other 
mechanisms related to these, or other impurities in silicon, and of their interaction with 
the vacancy and/or interstitial. The density of the reverse current in the detector after 
irradiation is estimated. Comparison with experimental published data of the time 
evolution of the concentration of defects is performed, as well as with measurements of 
the density of the leakage current. For different discrepancies, some explanations are 
suggested. A special application considered in the paper is the simulated case of the 
behaviour of silicon detectors operating in the pion field simulated for the future 
conditions at the new LHC (Large Hadron Collider) accelerator.

\section{Production of primary defects}
A point defect in a crystal is an entity that causes an interruption in the lattice 
periodicity. In this paper, the terminology and definitions in agreement with M. Lannoo 
and J. Bourgoin \cite{1} are used in relation to defects.

The basic assumption of the present model is that vacancies and interstitials are produced 
in materials exposed to radiation in equal quantities, uniformly in the bulk of the 
sample. They are the primary radiation defects, being produced either by the incoming 
particle, or as a consequence of the subsequent collisions of the primary recoil in the 
lattice.

The concentration of the primary radiation induced defects per unit fluence (CPD) in the 
semiconductor material has been calculated using the explicit formula (see details, e.g. 
in references \cite{2,3}):
\begin{equation}
CPD	\left(E\right)= \frac{N_{Si}}{2E_{Si}} \int \sum _{i} \left( \frac{d\sigma}{d\Omega} 
\right)_{i,Si} L(E_{Ri})_{Si} d\Omega=\frac{1}{N_A} \frac{N_{Si}A_{Si}}{2E_{Si}} 
NIEL\left(E\right)
\end{equation}
where $E$ is the kinetic energy of the incident particle, $N_{Si}$ is the atomic density 
in silicon, $A_{Si}$ is the silicon atomic number, $E_{Si}$ - the average threshold energy 
for displacements in the semiconductor, $E_{Ri}$  - the recoil energy of the residual 
nucleus produced in interaction $i$, $L(E_{Ri})$ - the Lindhard factor that describes the 
partition of the recoil energy between ionisation and displacements and 
$(d\sigma/d\Omega)_i$ - the differential cross section of the interaction between the 
incident particle and the nucleus of the lattice for the process or mechanism $i$, 
responsible in defect production. $N_A$ is Avogadro's number. The formula gives also the 
relation with the non ionising energy loss ($NIEL$). It is important to observe that there 
exists a proportionality between the $CPD$ and $NIEL$ only for monoelement materials.

For $CPD$ produced by pions, the pion - silicon interaction has been modelled \cite{4} and 
the energy dependencies of the Lindhard factors have been calculated in the frame of 
analytical approximations for different recoils in Si \cite{5}.

The concentration of primary defects produced by protons, neutrons, electrons and photons 
have been obtained from the $NIEL$. The calculations of Summers and co-workers for proton 
and electron $NIEL$ in silicon from reference \cite{6}, the calculations of proton, 
electron and photon $NIEL$ of Van Ginneken \cite{7} as well as those of Ougouang for 
neutrons \cite{8} have been considered.

\begin{figure}[ht]
\centering
\includegraphics[width=.8\textwidth, clip, angle=90]{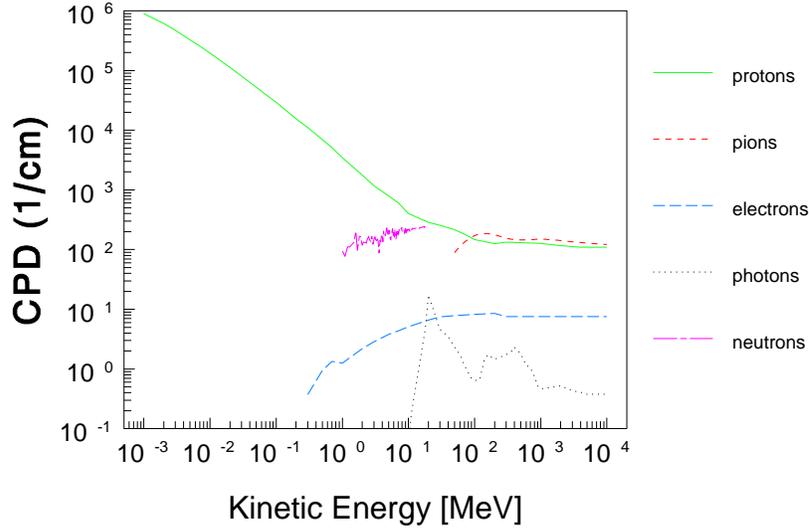}
\caption{\small{Energy dependence of the concentration of primary defects on unit fluence 
induced by protons, pions, electrons, photons and neutrons in silicon - see text for 
details.}}
\label{f1}
\end{figure}

In Figure \ref{f1}, the dependence of the $CPD$ on the particle kinetic energy is 
presented: for pions, our calculations from reference \cite{9} have been used; for 
protons, in the energy range $10^{-3} \div 1$ MeV the calculations of Summers, in the 
range $1 \div 200$ MeV the average between those of Summers and of Van Ginneken, while in 
the range $200 \div 10000$ MeV, Van Ginneken's. The curve for electrons uses, up to 1 MeV 
only the values from reference \cite{6}, in the range $1 \div 200$ MeV an average between 
the values from references \cite{6} and \cite{7}, and after 200 MeV only from \cite{7}. 
The curves for photons and neutrons are calculated from Van Ginneken's \cite{7} and 
Ougouang's \cite{8} respectively.

The main source of errors in the calculated concentration of defects comes from the 
modelling of the  particle - nucleus interaction and from the number and quality of the 
experimental data available for these processes. 
Due to the important weight of annealing processes, as well as to their very short time 
scale, CPD is not a measurable physical quantity.

In silicon, vacancies and interstitials are essentially unstable and interact via 
migration, recombination, annihilation or produce other defects. 

\section{The kinetics of radiation induced defects}

In the frame of the model, equal concentrations of vacancies and interstitials are 
supposed to be produced by irradiation, in much greater concentrations than the 
corresponding thermal equilibrium values, characteristic to each temperature. Both the 
pre-existing defects and those produced by irradiation, as well as the impurities, are 
assumed to be randomly distributed in the solid. An important part of the vacancies and 
interstitials annihilate. The sample contains certain concentrations of impurities which 
can trap interstitials and vacancies respectively, and form stable defects.

In the present paper, vacancy-interstitial annihilation, interstitial migration to sinks, 
divacancy and vacancy impurity complex formation  ($VP$, $VO$, $V_2O$), are considered. 
The mechanisms of formation of higher order defects involving vacancy and oxygen can be 
added, as well as the effects of other impurities, e.g. carbon. 

This picture could be described in terms of chemical reactions by the kinetic scheme:
\begin{equation}                                                        
V+I\ _{\overleftarrow{K_1}} ^{\underrightarrow{G}}\text{annihilation}
\end{equation}
\begin{equation}                                                       
I\stackrel{K_2}{\rightarrow } \text{sinks}
\end{equation}
\begin{equation}							
V+P\ _{\overleftarrow{K_5}} ^{\underrightarrow{K_5}}\ VP
\end{equation}
$VP$ is the $E$ centre.
\begin{equation}							
V+O\ _{\overleftarrow{K_3}} ^{\underrightarrow{K_4}}\ VO
\end{equation}
$VO$ is the $A$ centre.
\begin{equation}						
V+V\ _{\overleftarrow{K_8}} ^{\underrightarrow{K_7}}\ V_2
\end{equation}
\begin{equation}						
V+A\ _{\overleftarrow{K_{10}}} ^{\underrightarrow{K_{9}}}\ V_2O
\end{equation}

The bimolecular recombination law of interstitials and vacancies is supposed to be a valid 
approximation for the present discussion, because at the concentrations of vacancies of 
interest, only a small fraction of defects anneals by correlated annihilation if their 
distribution is random (see the discussion in reference \cite{10}).

The multivacancy oxygen defects as, e.g. $V_3O$, $V_2O_2$, $V_3O_2$, $V_3O_3$, are not 
considered in the model.

The reaction constant $K_1$ (corresponding to vacancy - interstitial annihilation) is 
determined by the diffusion coefficient of the interstitial atom to a substitutional trap:
\begin{equation}
K_1=30\nu \exp \left( -E_{i1}/k_BT\right)
\end{equation}
where $E_{i1}$ is the activation energy of interstitial migration and $\nu$ 
the vibrational frequency. The reaction constant in process (2) is 
proportional to the sink concentration $\alpha$:

\begin{equation}
K_2=\alpha \nu \lambda ^2\exp \left( -E_{i1}/k_BT\right)                                                 
\end{equation}
with $\lambda$ the jump distance.

Lee and Corbett \cite{11} argue that divacancies, vacancy-oxygen and divacancy-oxygen 
centres are equally probable below 350 $^o$C; thus, $K_3$, $K_5$, $K_7$ and $K_9$, that 
describe the formation of vacancy - impurity complexes and of divacancies, are determined 
by the activation energy of vacancy migration, $E_{i2}$, and are given by:
\begin{equation}
K_3=K_5=K_7=K_9=30\nu \exp \left( -E_{i2}/k_BT\right)
\end{equation}
while $K_4$, $K_6$, $K_8$ and $K_{10}$ are related to the activation energies of 
dissociation of the $A$, $E$, $V_2$ and $V_2O$ centres respectively.

\begin{equation}
K_4=5\nu \exp \left( -E_A/k_BT \right)					
\end{equation}
\begin{equation}
K_6=5\nu \exp \left( -E_E/k_BT \right)					
\end{equation}
\begin{equation}
K_8=5\nu \exp \left( -E_{V_2)}/{k_BT}\right)				
\end{equation}
\begin{equation}
K_{10}=5\nu \exp \left( -E_{V_2O}/{k_BT}\right)				
\end{equation}
where $E_A$, $E_E$, $E_{V_2}$ and $E_{V_2O}$ are the dissociation energies of the $A$, 
$E$, $V_2$ and $V_2O$ complexes respectively.

$G$ is the generation rate of vacancy-interstitial pairs, and is given by the product of 
$CPD$ by the irradiation flux. Thermal generation is neglected, this approximation 
corresponding to high irradiation fluxes. 

In the simplifying hypothesis of random distribution of $CPD$ for all particles, two 
different particles can produce the same generation rate for vacancy-interstitial pairs.

\begin{equation}
G=[\left(CPD\right)_{part.a}\left(E_1)\right]\cdot\Phi_{part.a}(E_1)=[\left(CPD\right)_{pa
rt.b}\left(E_2)\right]\cdot\Phi_{part.2(E_2)}
\end{equation}
is fulfilled.

Here, $\Phi$ is the flux of particles ($a$) and ($b$) respectively, and $E_1$ and $E_2$ 
their corresponding kinetic energies.
The system of coupled differential equations corresponding to the reaction scheme (2)-(7) 
cannot be solved analytically. 

The following values of the parameters have been used: $E_{i1}$ = 0.4 eV, $E_{i2}$ = 0.8 
eV, $E_A$ = 1.4 eV, $E_B$ = 1.1 eV, $E_{V_2}$ = 1.3 eV, $E_{V_2}$ = 1.6 eV, $E_{V_2O}$ = 
$10^{13}$ Hz, $\lambda$ = $10^{15}$ cm$^2$, $\alpha = 10^{10}$ cm$^{-2}$.

Defect concentrations, as well as their time evolution, have been calculated solving 
numerically the system of coupled differential equations.

We would like to underline the specific importance of the irradiation and annealing 
history (initial material parameters, type of irradiation particles, energetic source 
spectra, flux, irradiation temperature, measurement temperature, temperature and time 
between irradiation and measurement) on defect evolution.

\begin{figure}[ht]
\centering
\includegraphics[width=.8\textwidth, clip, angle=90]{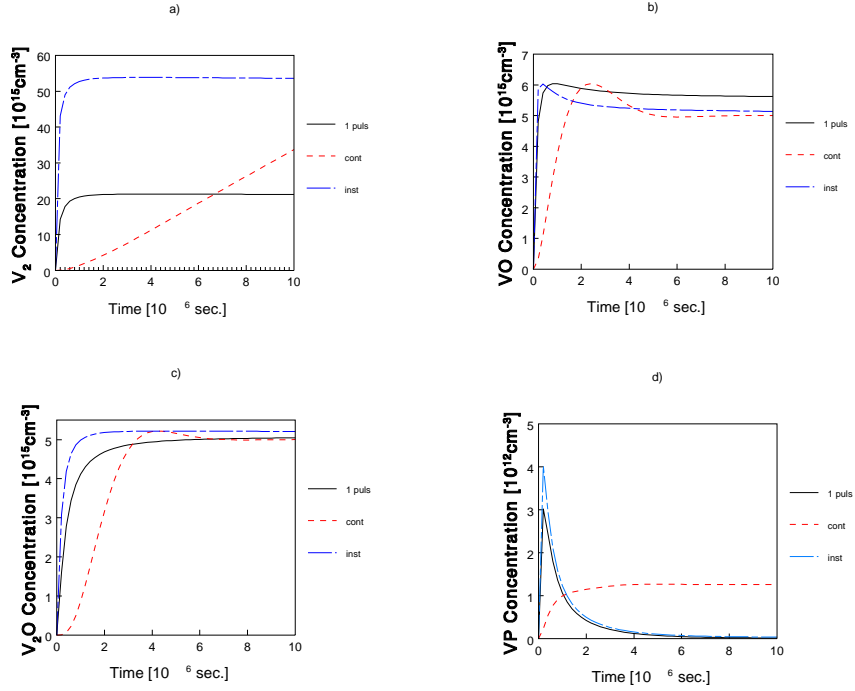}
\caption{\small{Time dependence of the concentrations of: a)$V_2$, b)$VO$, c)$V_2O$ and 
d)$VP$,
induced in silicon with 10$^{14}$ P/cm$^3$ and 5x10$^{16}$ O/cm$^3$, irradiated with 200 
MeV kinetic energy pions, at a total fluence of $10^{15}$ pions/cm$^2$ in different 
irradiation conditions - see text.}}
\label{f2}
\end{figure}

In Figures 2a $\div$ d, the formation and time evolution of the divacancy, vacancy-oxygen, 
divacancy-oxygen, and vacancy-phosphorous is modelled in silicon containing the initial 
concentrations of impurities: 10$^{14}$ P/cm$^3$   and 5x10$^{16}$ O/cm$^3$, and 
irradiated with pions with about 200 MeV kinetic energy (corresponding to the in their 
maximum of $CPD$ in the energetic distribution), at a total fluence of 10$^{15}$ 
pions/cm$^2$, in different irradiation conditions. The instantaneous irradiation process 
is an ideal case \cite{12} where the total fluence is received by the material at time t = 
0, and only the relaxation process is studied. This process supposes that the annealing 
effects are not present during irradiation. The second case considered is: the irradiation 
is performed in a single pulse for a time 2x10$^4$ seconds, followed by relaxation, and 
the third case is a continuum irradiation process with a generation rate of defects equal 
with 5x10$^{10}$ pions/cm$^2$/s. In these last two cases, annealing takes place during 
irradiation. During and after irradiation, the temperature is kept at 293K.
As expected, after instantaneous irradiation the concentrations of defects are higher in 
respect with "gradual" irradiation. A special attention must be paid to the formation of 
$VO$ and $V_2O$ centres, for which the same concentration is attained both in the case of 
continuous and instantaneous irradiation.

\begin{figure}[ht]
\centering
\includegraphics[width=.8\textwidth]{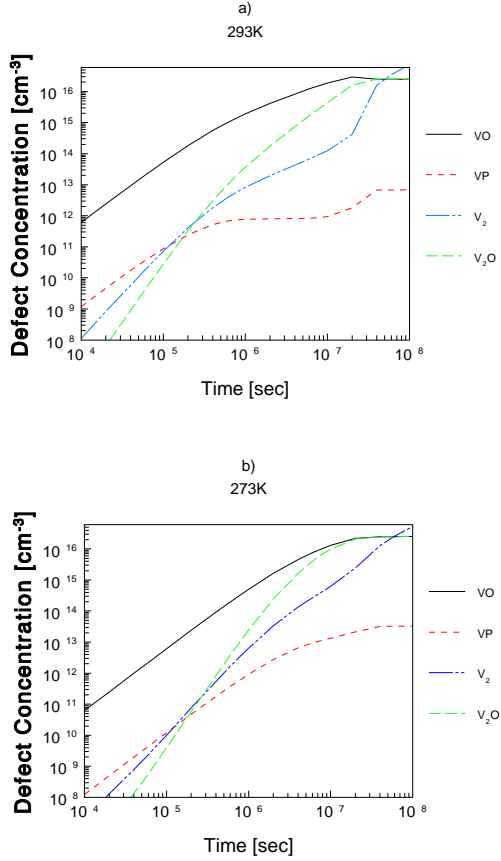}
\caption{\small{Time dependence of the concentrations of $VO$, $VP$, $V_2$ and $V_2O$  
induced in silicon with $10^{14}$ P/cm$^3$ and 5x10$^{16}$ O/cm$^3$, irradiated with 200 
MeV kinetic energy pions a total fluence of $10^{15}$ pions/cm$^2$ with the flux estimated 
for LHC, for 293 and 273 K.}}
\label{f3}
\end{figure}

The effect of the decrease of temperature, from 293 to 273 K during irradiation and 
annealing, is presented in Figures 3a and b. The material contains the same phosphorous 
and oxygen concentrations as in the modelled case presented in Figure 2, and was 
irradiated with pions of 200 MeV kinetic energy, receiving continuously a fluence of 
$10^{15}$ pions/cm$^2$ in ten years, in accord to the pions simulated radiation field at 
LHC \cite{13,14}.

The increase of temperature increases the rate of all defect formation. In the case of 
$VO$ and $VP$ a plateau in the time dependencies is attained. Only for $VP$  the plateau 
value is temperature sensitive.

\begin{figure}[ht]
\centering
\includegraphics[width=.8\textwidth]{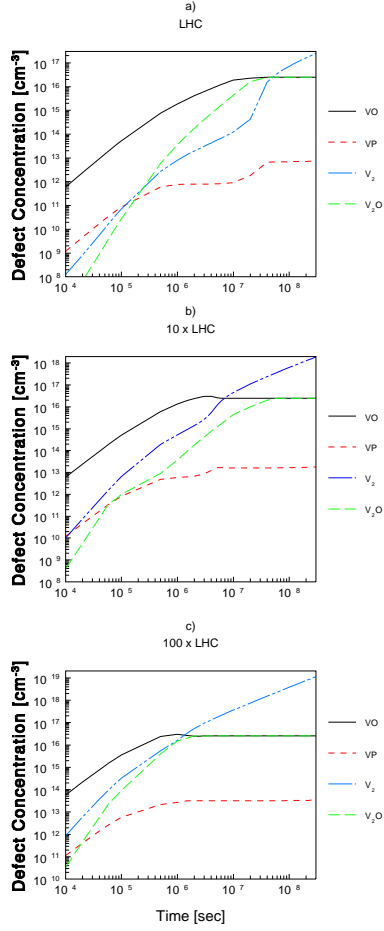}
\caption{\small{Time dependence of the concentrations of  $VO$, $VP$, $V_2$ and $V_2O$  
induced in 
silicon with $10^{14}$ P/cm$^3$ and $5×10^{16}$ O/cm$^3$, by continuous irradiation with 
200 MeV kinetic energy pions with the flux: :a) estimated for LHC, b) 10 times the flux 
estimated for LHC, c) 100 times the flux estimated for LHC, at 293 K.}}
\label{f4}
\end{figure}

The rate of generation of primary defects in silicon influences the concentrations of all 
stable defects. In Figures 4a $\div$ c, the time evolution of the concentrations of $VO$, 
$VP$, $V_2$ and $V_2O$ in silicon with 10$^{14}$ P/cm$^3$ and 5x10$^{16}$ O/cm$^3$, 
irradiated with pions of 200 MeV with the flux estimated for $LHC$, 10 and 100 times 
higher respectively, is presented. It can be observed that at the $LHC$ generation rate 
for $CPD$, after 3x10$^7$ seconds an equilibrium is established for the $VP$ complex: its 
rate of formation equals its rate of dissociation - Fig. 4a. This time is shorter for 
higher generation rate, as can be observed in Figures 4b and 4c. The value of the plateau 
concentration for the vacancy-oxygen complex is attained after around the same time as the 
plateau for the  concentration, in conditions of the $LHC$ generation rate, and a shorter 
time, about 2x10$^6$ sec. for a rate hundred times higher than the $LHC$ one. For other 
defects, as divacancies and divacancy-oxygen, the processes to established the equilibrium 
are very slow.

The formation of divacancy-oxygen is delayed in respect to vacancy oxygen, and for long 
exposure times, the same value for the concentration is obtained.

The effect of oxygen in irradiated silicon has been a subject of intensive studies in 
remote past. In the last decade a lot of studies have been performed to investigate the 
influence of different impurities, especially oxygen and carbon, as possible ways to 
enhance the radiation hardness of silicon for detectors in the future generation of 
experiments in high energy physics - see, e.g. references \cite{15,16}. These impurities 
added to the silicon bulk modify the formation of electrically active defects, thus 
controlling the macroscopic device parameters. If silicon is enriched in oxygen, the 
capture of radiation-generated vacancies is favoured by the production of the 
pseudo-acceptor complex vacancy-oxygen. Interstitial oxygen acts as a sink of vacancies, 
thus reducing the probability of formation of the divacancy related complexes, associated 
with deeper levels inside the gap. For this purpose, in the model, the effects of the 
initial oxygen concentration in silicon was studied. In Figures 5 a, b, c, d the time 
dependencies of $V_2$, $VO$, $V_2O$ and $VP$ are presented, for silicon containing  
10$^{15}$, 10$^{16}$, 10$^{17}$, and 10$^{18}$ atoms/cm$^3$ initial oxygen concentrations.

\begin{figure}[ht]
\centering
\includegraphics[width=.8\textwidth,, clip, angle=90]{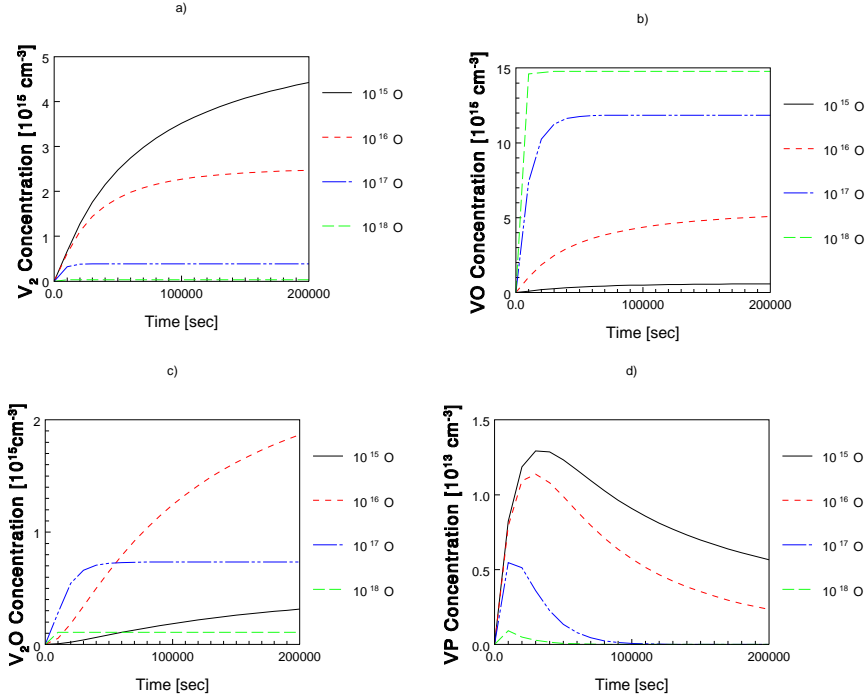}
\caption{\small{Effect of oxygen doping concentration on the time dependence of the 
concentrations of: a)$V_2$, b)$VO$, c)$V_2O$ and d)$VP$, induced in silicon with $10^{14}$ 
P/cm$^3$ irradiated with 200 MeV kinetic energy pions at total fluence of 10$^{14}$ 
pions/cm$^2$ in one pulse.}}
\label{f5}
\end{figure}

One can observe that vacancy-oxygen formation in oxygen enriched silicon is favoured in 
respect to the generation of $V_2$, $V_2O$ and $VP$, confirming the considered hypothesis, 
so, for detector applications the leakage current is decreased. At high oxygen 
concentrations, the concentration of $VO$ centres saturates starting from low fluences.

A difficulty in the comparison of model predictions with experimental data is the 
insufficient information in published papers regarding the characterisation of silicon, 
and on the irradiation (flux, temperature during irradiation and measurement, irradiation 
time, time and temperature between irradiation and measurement) for most of the data.

\begin{figure}[ht]
\centering
\includegraphics[width=.8\textwidth]{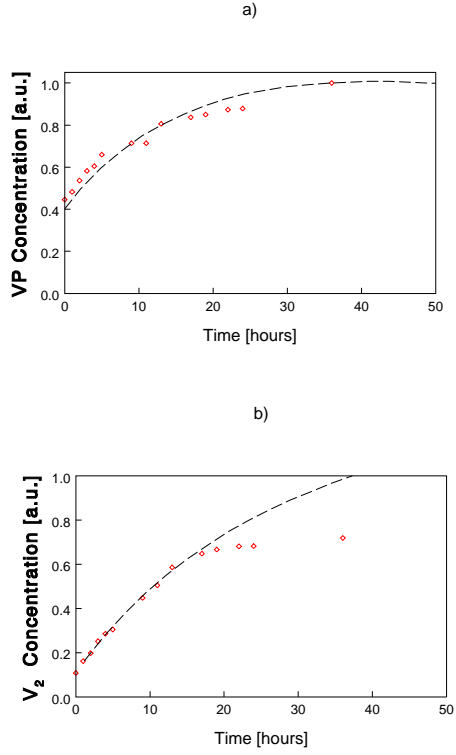}
\caption{\small{Time dependence of a)$VP$ and b)$V_2$ concentrations after electron 
irradiation:
points: experimental data from reference \cite {17} and dashed line: model calculations.}}
\label{f6}
\end{figure}

For electron irradiation, our simulations are in agreement with the measurements presented 
in reference \cite{17}, where defect concentrations are presented as a function of the 
time after irradiation. In Figure 6, both measured and calculated dependencies of the $VP$ 
and $V_2$ concentrations are given. The irradiation was performed with 2.5 MeV electrons, 
up to a fluence of 10$^{16}$ cm$^{-2}$. A good agreement can be observed for the 
concentration of $VP$, see Figure 6a, while for the divacancy, Figure 6b, the experimental 
data attain a plateau faster, and at smaller values than the calculations. The relative 
values are imposed by the arbitrary units of experimental data.

A good agreement has also been obtained for hadron irradiation. For example, the sum of 
the calculated $VP$ and $V_2$ concentrations (8x10$^{12}$ cm$^{-3}$) induced in silicon by 
5.67x10$^{13}$ cm$^{-2}$ 1 MeV neutrons, are in accord with the experimental value of 
11.2x10$^{12}$ cm$^{-3}$ reported in reference \cite{18}.

\section{Correlation with detector parameters}
It is well known that the dark current in a $p-n$  junction is composed by three different 
terms: the diffusion current, caused by the diffusion of the minority charge carriers 
inside the depleted region; the generation current, created by the presence of lattice 
defects inside the bulk of the detector; and surface and perimetral currents, dependent on 
the environmental conditions of the surface and the perimeter of the diode. The appearance 
of the defects after irradiation corresponds therefore in an increase of the leakage 
current of the detector by its generational term.

Inside the depleted zone, $n, p << n_i$ ($n_i$ is the intrinsic free carrier 
concentration), each defect with a bulk concentration $N_T$ causes a generation current 
per unit of volume of the form \cite{19}:

\begin{equation}
I=qU=q<v_t>n_t \frac{\sigma_n \sigma_p N_T}{\sigma_n \gamma_n 
e^{\left(E_t-E_i\right)/k_BT}+\sigma_p \gamma_p e^{\left(E_i-E_t\right)/k_BT}}
\end{equation}
where $\gamma _n$ and $gamma _{p}$ are degeneration factors, $\sigma _n$ ($\sigma _p$) are 
the cross sections for majority (minority) carriers of the trap, $E_i=(E_C-E_V)/2$ and 
$<v_t>$ is the average between electron and hole thermal velocities.
In the case of $E$ and $A$ centres and $V_2^-$  and $V_2^(--)$  defects, the current 
concentration can be expressed in the simple form:
\begin{equation}
I=qU=q<v_t>n_t \frac{\sigma_n}{\sigma_p}N_Te^{\left(E_t-E_i\right)/k_BT}
\end{equation}

The primary effect in the recombination process is the change the charge state of the 
defect. The different charge states of the same deep centre may have different barriers 
for migration or for reacting with other centres. Thus, carrier capture can either enhance 
or retard defect migration or particular defect reactions. As a characteristics for 
detectors (as diode junction), the defect kinetics is dependent to the reverse - bias 
voltage during the irradiation \cite{20}.

The comparison between theoretical and experimental generation current densities after 
irradiation shows a general accord between experiment and the model results for the lepton 
irradiation and large discrepancies for the hadron case.

There could be several reasons for the observed discrepancies.

The model hypothesis of defects distributed randomly in semiconductors exclude the 
possibility of cluster defects. For this case, other mechanisms of defect formation are 
necessary, which suppose different reaction rates and correlation between the constituent 
defects of the cluster.

In the Shockley-Read-Hall model used for the calculation of the reverse current, each 
defect has one level in the gap, and the defect levels are uncoupled, thus the current is 
simply the sum of the contributions of different defects. In fact, the defects could have 
more levels, and charge states, as is the case of the divacancy, and  also could be 
coupled, as in the case of clusters. As shown in the literature \cite{21,22}, both cases 
can produces modifications of the generation rate.

Also the multivacancy oxygen defects as, e.g. $V_3O$, $V_2O_2$,  $V_3O_2$,  $V_3O_3$, are 
not considered in the model.

\begin{figure}[ht]
\centering
\includegraphics[width=.8\textwidth, clip, angle=90]{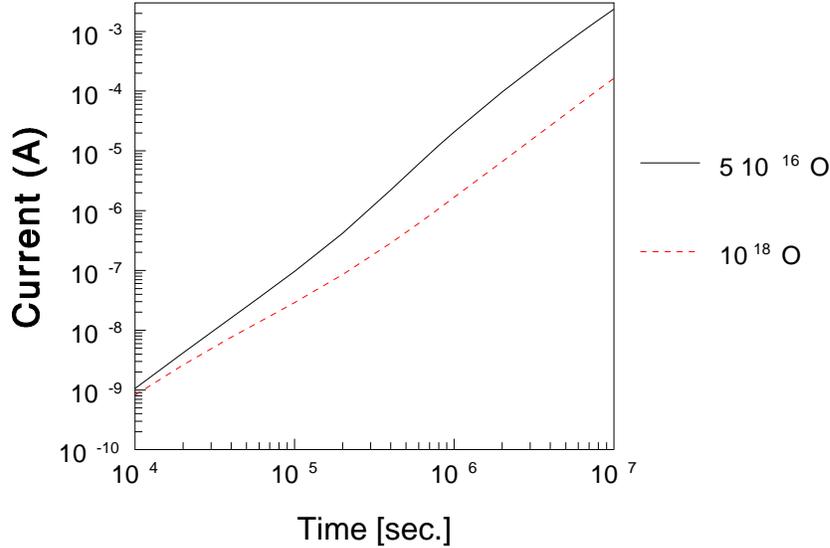}
\caption{\small{Time dependence of the reverse current after 200 MeV kinetic energy pions 
irradiation, with the rate estimated for LHC, at 293K, for silicon containing: 
a)5x10$^{16}$ cm$^{-3}$ and b)$10^{16}$ cm$^{-3}$ oxygen.}}
\label{f7}
\end{figure}

A model estimation of the time dependence of the leakage current, in conditions of 
continuous irradiations with pions of 200 MeV kinetic energy, in the conditions of the 
$LHC$ \cite{13,14} and at 293K is presented in Figure 7, for two concentrations of oxygen 
in silicon: 5x10$^{16}$ cm$^{-3}$ and 10$^{16}$ cm$^{-3}$ respectively. As underlined 
before, oxygen incorporation in silicon has beneficial effects, decreasing the reverse 
current. This conclusion is valid in the hypothesis of random distribution of defects 
inside the depleted zone of the p-n junction. These values are probably underestimated.

\section{Summary}
A phenomenological model that describes silicon degradation due to irradiation, the 
kinetics of defects toward equilibrium, and the influence on the reverse current of 
detectors was developed. 

The production of primary defects (vacancies and interstitials) in the silicon bulk was 
considered in the frame of the Lindhard theory, and considering the peculiarities of the 
particle - silicon nuclei interaction.

The mechanisms of formation of stable defects and their evolution toward equilibrium was 
modelled, and the concentrations of defects were calculated solving numerically the system 
of coupled differential equations for these processes. Vacancy-interstitial annihilation, 
interstitial migration to sinks, vacancy-impurities complexes ($VP$, $VO$ and $V_2O$), and 
divacancy formation were considered in different irradiation conditions, for different 
concentrations of impurities in the initial semiconductor material and at different 
temperatures of irradiation. The calculated results suggest the importance of the 
conditions of irradiation, temperature and annealing history. The model supports the 
experimental studies performed to investigate the influence of oxygen in the enhancement 
of the radiation hardness of silicon for detectors. The $VO$ defects in oxygen enriched 
silicon is favoured in respect to the other stable defects, so, for detector applications 
it is expected that the leakage current decreases after irradiation. The second result in 
the model is that at high oxygen concentrations, this defect saturates starting from low 
fluences. 

Most of the model calculations simulates some of the pion field estimated at the new LHC 
accelerator, where the silicon detector will operate under continuum generation and 
annealing.

The density of the reverse current in detectors after irradiation is estimated, compared 
with experimental available data and for discrepancies some explanations are suggested.

\section{Acknowledgements}

The authors are very grateful to Professor Gh. Ciobanu from the Bucharest University for 
helpful discussions during the course of this work.

\end{document}